# Learning-Inspired Fuzzy Logic Algorithms for Enhanced Control of Oscillatory Systems

1st Vuong Anh Trung
*Faculty of Aviation Technology*
Air Defence – Air Force Academy
Hanoi, Vietnam
vuonganhtrung@gmail.com

2nd Thanh Son Pham
Faculty of Computing,
FPT University, FPT Greenwich Centre
Danang, 550000, Vietnam
email: sonpt54@fe.edu.vn

3rd Truc Thanh Tran
Faculty of Computing,
FPT University, FPT Greenwich Centre
Danang, 550000,Vietnam
ORCID: https://orcid.org/0000-0001-9186-7504 ; email: tructt16@fe.edu.vn

4th Tran Le Thang Dong
*The Center of Electrical Engineering*
Duy Tan University
Danang, 550000, Vietnam
tranthangdong@duytan.edu.vn

5th Tran Thuan Hoang
*The Center of Electrical Engineering* Duy Tan University
Danang, 550000, Vietnam
tranthuanhoang@duytan.edu.vn

*Abstract* - **The transportation of sensitive equipment often suffers from vibrations caused by terrain, weather, and motion speed, leading to inefficiencies and potential damage. To address this challenge, this paper explores an intelligent control framework leveraging fuzzy logic, a foundational AI technique, to suppress oscillations in suspension systems. Inspired by learning-based methodologies, the proposed approach utilizes fuzzy inference and Gaussian membership functions to emulate adaptive, human-like decision-making. By minimizing the need for explicit mathematical models, the method demonstrates robustness in both linear and nonlinear systems. Experimental validation highlights the controller's ability to adapt to varying suspension lengths, reducing oscillation amplitudes and improving stability under dynamic conditions. This research bridges the gap between traditional control systems and learning-inspired techniques, offering a scalable, data-efficient solution for modern transportation challenges.**

## I. INTRODUCTION

During transportation, equipment is often subjected to vibrations caused by terrain, weather, and motion speed, leading to system oscillations that can cause damage and reduce operational efficiency. Minimizing and eliminating these vibrations is a critical challenge in ensuring system reliability and safety. While various input determination algorithms have been widely used for vibration control, they are primarily effective for linear control systems. For nonlinear systems, a more adaptable and efficient control approach is essential [1, 2].

Currently, in the process of transporting military equipment for the Air Defense-Air Force, specifically in the process of transporting missile ammunition on specialized vehicles (TZM, ...) [3], automatic crane systems in production lines, and maintenance of rocket ammunition [4]. There is a phenomenon of shaking (oscillation) leading to incorrect parameters. This oscillation reduces the efficiency of transport and operation of the conveyed gas and poses a danger to people. The urgent requirement is to minimize and eliminate those vibrations (oscillations) [5]. Many studies give the input determination algorithm widely used to control the transport of devices without clear oscillation feedback. Based on linear control theories and optimal control theories, the research proposes algorithms to minimize oscillations. However, these algorithms are only practical when used for linear control systems. For nonlinear systems, developing a more suitable control algorithm is necessary. Currently, the fuzzy control system has solved nonlinear problems through experience and practical characteristics of the system. The content of the article proposes to use fuzzy logic algorithms to suppress oscillations effectively, with high accuracy and ease to apply in practice to both linear and nonlinear control systems [6,7,8].

## II. FUZZY LOGIC

There are different ways to use a fuzzy controller in the structure of an automatic control system. The operation of the fuzzy controller depends on the method of drawing human conclusions, which are then installed on the computer-based on fuzzy logic. A fuzzy controller comprises three basic blocks: a fuzzy block, a composing device, and a diversifying block. There is also an input and output interface block (Figure 1).

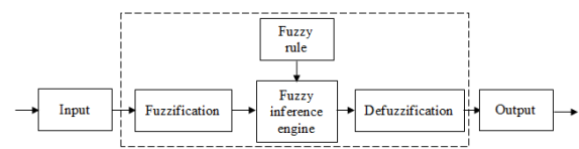

Fig 1. General fuzzy controller

The principle of synthesizing a fuzzy controller is wholly based on mathematical methods based on the definition of input/output linguistic variables and the choice of control rules. Because fuzzy controllers can handle input/output values with high accuracy, they meet the control problem entirely [9,10,11].

A fuzzy controller is used for both the position control system and the vibration suppression control system to suppress oscillations in lifting devices. Dividing the control task into oscillation suppression and position control allows the individual controller to be tailored to specific requirements. The content of the article only studies the control task of oscillation suppression [12,13,14].

## III. THE CONTROL SYSTEM WITH FUZZY CONTROLLER

The block diagram of the control system with a fuzzy controller is shown in Figure 2. The control algorithm is implemented through 3 stages. In the first stage, the fuzzy block transforms the input physical variables $(x_1, x_2, ..., x_n)$. Use the composition rules to determine the terms Bj of the output variable. The fuzzified transforms the fuzzy sets (element $B_j$) into the exact value of the physical variable $y$ (control signal).

The method uses a fuzzy controller to suppress oscillations by using the deviation angle signals and the deviation angle's derivative. At the same time, these signals must ensure the control quality at an acceptable level and not cause noise [15,16]. The rule base (logic law) of the fuzzy controller is based on the implementation of the following principles:

- If the load-deflection marks and the load-deflection angular speed are the same, the load will move out of the equilibrium position, and a control action of the same sign must be introduced (i.e., moving the mechanism in the deflection direction). of the load);
- If the load-deflection mark and the load-deflection angular speed are opposite, a zero or small control action shall be provided to reduce the deflection;
- If the load-deflection angle and the load-deflection angular speed are zero, the load is in equilibrium, and no control action is required.

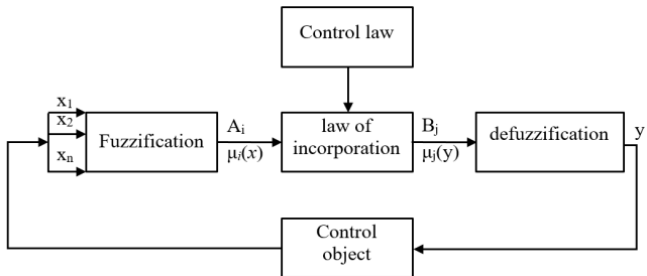

Fig 2. Block diagram of the control system using fuzzy controller

When using fuzzy control to suppress oscillation using 5 linguistic variables for input Ai={large negative, medium negative, zero, medium positive, large positive}, ($(A_i = \{NB, NM, Z, PM, PB\})$) and 7 output language variables Bj={large negative, medium negative, small negative, zero, small positive, medium positive, large positive}, ($B_j = \{NB, NM, NS, Z, PS, PM, PB\}$), the elements are arranged in order of increasing amplitude of the control signal, taking into account the sign ($NB$ corresponding control variable allows maximum negative effect, corresponding PB control variable allows maximum negative effect). Maximum positive dynamics, the larger the linguistic variables, the higher the accuracy, but the number of composition rules will increase. Mamdani and Takagi-Sugeno algorithms are used to infer fuzzy logic when implementing control by determining the center of gravity.

In the control system in Figure 2, the fuzzy controller has 2 input variables $x_1 = \varphi$ and derivative and output variable $y = -u$, where is the load deflection angle, $u$ is the driving characteristic of the device hang. Using 3 fuzzy sets for each input variable, the fuzzy block $A_1 = \{N\varphi\ Z\varphi\ P\varphi\}$, and $A_2 = \{N\Phi\ Z\Phi\ P\Phi\}$, the defuzzification block also uses three fuzzy sets $Y = \{Nu\ Zu\ Pu\}$. To blur and defuzzify in fuzzy controller using Gaussian function [17,18]:

$$\mu_k(z) = e^{-\frac{(z-c_k)^2}{2(\sigma_k)^2}} \quad (1)$$

Where: $z$ - input or output variable; $c_k$ - maximum coordinates of the $k_{th}$ membership function along the z-axis; $\sigma_k$ - coefficient of the kth membership function.

The coordinates $c_k$ are evenly distributed over the range of values of the physical variable: $c_1 = z_{min}$, $c_2 = 0$, $c_3 = z_{max}, z_{min} = z_{max}, k = \{1,3\}$.

The coefficient σk is chosen so that:

$$\mu_k\left(\frac{c_k + c_{k+1}}{2}\right) = \frac{1}{2} \quad (2)$$

From (1), (2) deduce:

$$\sigma_k = \sqrt{\frac{c_{k+1} - c_k}{8 \ln(2)}} \quad (3)$$

The values of $\varphi_{min}$ and $\varphi_{max}$ are determined according to the requirements of the maximum and minimum permissible load deviation [19,20]. From that, the following composition rule table can be built:

TABLE I. LAW OF COMPOSITION OF LOAD DEFLECTION ANGLE AND RATE OF CHANGE (DERIVATIVE) OF LOAD DEFLECTION ANGLE

|    | NB | NM | Z  | PM | PB |
|----|----|----|----|----|----|
| NB | NB | NB | NM | NS | Z  |
| NM | NB | NM | NS | Z  | PS |
| Z  | NM | NS | Z  | PS | PM |
| PM | NS | Z  | PS | PM | PB |
| PB | Z  | PS | PM | PB | PB |

The rule base and defuzzification process are described in Table I, assuming the maximum values $\varphi_{max}, \dot{\varphi}_{min}, u_{max}$ are taken to be 1 relative unit (1 o.e.). Figure 3 is the result of the composition rule using the rules and the defuzzification process in the state $\varphi = -0.2\ o.e.$, $u = 0.6\ o.e.$ now $u = 0.193\ o.e.$ [21,22,23].

The result of performing fuzzy logic inference (getting a conclusion in the form of a fuzzy set corresponding to the current values of the input, on the basis of using fuzzy rules and operations): Output terms: are assigned values of functions of input variables with the general condition $\mu(y) \leq \mu_i(x)$, i.e. according to the principle of least (association operator) $\mu(y) = \min(\mu k_1(\varphi), \mu k_2(\dot{\varphi}))$, where $k_1$ and $k_2$ are the number of membership functions, respectively [24,25].

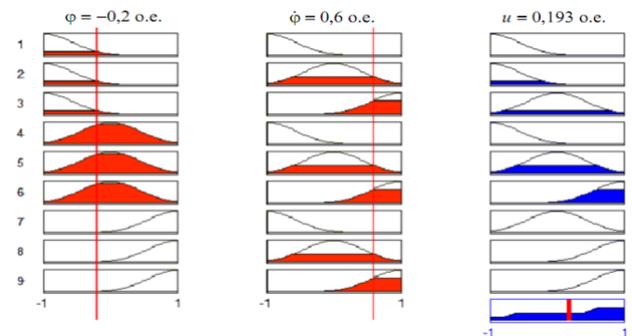

Fig 3. Representation of the result of defuzzification composition rule using Matlab

The characteristic of the fuzzy controller is developed, describing the change of control operation depending on the load deflection angle and its derivative has the form of Fig 4.

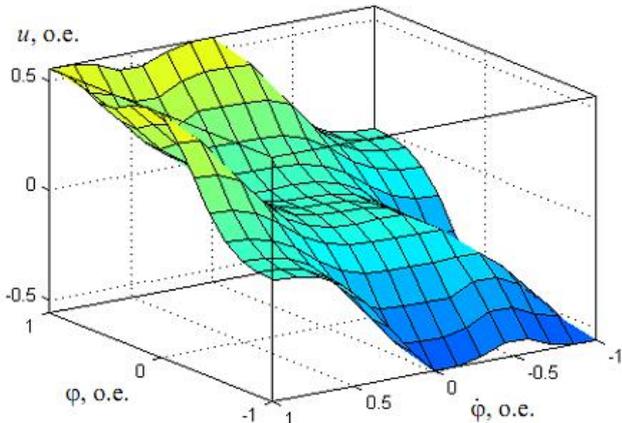

Fig 4. Fuzzy inference surface with three fuzzy sets for each variable using Matlab

In order to analyze the effectiveness of the oscillation suppression control system by fuzzy logic, it is necessary to study the sensitivity of the control system to changes in the length of the suspension. The results are taken as the suspension length changes up or down from the value used when setting the fuzzy controller. Figure 5 is the result obtained when changing the suspension length.

- L - actual value of the suspension length in relative units (o.e.);
- V - relative amplitude of load oscillation (which is the ratio between the amplitude of the load oscillation when using the fuzzy controller to the amplitude of the load oscillation when not using the fuzzy controller);
- $t_{\pi\pi}$ - oscillation time.

When $t_{\pi\pi} = 1 o.e.$ corresponding to the load oscillation period is $T_0 = 2\pi\sqrt{\frac{I_0}{g}}$, where $I_0$ is suspension length when using a dimming controller; $g$ is acceleration.

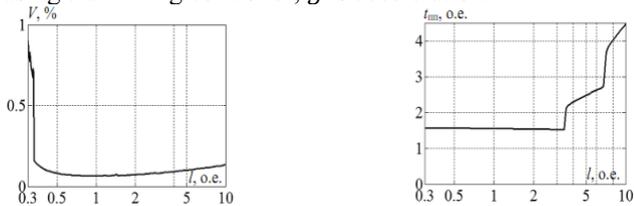

Fig 5. Sensitivity study results of the used system fuzzy controller when suspension length changes. a) The dependence of the amplitude of oscillation. b) The dependence of the oscillation time

From the above characteristics, it can be seen that when using the fuzzy controller, the amplitude of the oscillation of the suspension device will be greatly reduced when transported with a variable suspension length, the amplitude of oscillation increases sharply when the length of the suspension is changed. slings larger than $3.4 \, o.e$ (Fig 5a). When the length $l$ changes from $0.3 \, o.e.$ to about $3.4 \, o.e.$ then the oscillation amplitude is relatively stable, the time t$\pi\pi$ decreases from $1.57 \, o.e.$ to $1.52 \, o.e.$ and the time t$\pi\pi$ increases sharply when $l > 3.4 \, o.e$ (Fig 5b).

Through the analysis results, the nature of the stabilization process will increase when using a fuzzy controller, the stabilization process is characterized by the degree of oscillation suppression under different conditions. The advantage of using fuzzy controller to suppress oscillation of suspended equipment when transporting this case is when the input information is incomplete and unclear. However, the performance of the system depends on the rule bases for building composition rules and fuzzy control rules.

IV. CONCLUSION

To suppress the oscillations of the suspension devices during transportation, in this paper, it is proposed to use control systems using fuzzy logic algorithms that will ensure stability in a wide range (system length varies). change from 0.3 o.e to 3.4 o.e), the process of synthesizing the controller is simple, easy to apply in practice. The advantage of the fuzzy logic controller in this case is that it solves the problem quite like human thinking, does not require a mathematical model of the control object, so it can be applied to many complex systems, different, especially nonlinear systems.